\documentclass[11pt]{article}
\usepackage{BLOIS02,epsfig,latexsym,amssymb}
\def\Journal#1#2#3#4{{#1} {\bf #2}, #3 (#4)}

\def\NPB{{\em Nucl. Phys.} B}
\def\PLB{{\em Phys. Lett.}  B}
\def\PRL{\em Phys. Rev. Lett.}
\def\PRD{{\em Phys. Rev.} D}
\def\PR {{\em Phys. Rev.}}

\def\be{\begin{equation}}
\def\ee{\end{equation}}
\def\bea{\begin{eqnarray}}
\def\eea{\end{eqnarray}}

\newcommand{\bdi}{\begin{displaymath}}
\newcommand{\edi}{\end{displaymath}}
\newcommand{\bfi}{\begin{figure}}
\newcommand{\efi}{\end{figure}}

\newcommand{\beq}{\begin{equation}}
\newcommand{\eeq}{\end{equation}}
\newcommand{\beqa}{\begin{eqnarray}}
\newcommand{\eeqa}{\end{eqnarray}}
\newcommand{\no}{\nonumber}

\newcommand{\CS} {Chern--Si\-mons}

\newcommand {\YM}     {Yang--Mills}
\newcommand {\YMth}   {Yang--Mills theory}

\newcommand {\YMH}    {Yang--Mills--Higgs}

\newcommand {\rhs}    {right-hand side}
\newcommand {\sm}     {Standard Model}
\newcommand {\ew}     {electroweak}
\newcommand {\ewsm}   {electroweak Standard Model}
\newcommand {\bnv}    {baryon number violation}
\newcommand {\ewbnv}  {electroweak baryon number violation}
\newcommand {\gsim}{\mathrel{\hbox{\rlap{\lower.55ex \hbox {$\sim$}}
            \kern-.3em \raise.4ex \hbox{$>$}}}}
\newcommand {\lsim}{\mathrel{\hbox{\rlap{\lower.55ex \hbox {$\sim$}}
            \kern-.3em \raise.4ex \hbox{$<$}}}}
\def\id{\makebox[0.6ex][l]{$1$}{\rm l}}   

\begin{document}

\voffset=-5mm                                                 
\footskip=10mm                                                
\pagestyle{plain}                                             
\noindent  hep-ph/0209227            \hfill  KA--TP--11--2002 
\vspace*{-1\baselineskip}                                     

\vspace*{4cm}
\title{ELECTROWEAK BARYON NUMBER VIOLATION
       \footnote{to appear in the                      
       Proceedings of the XIV-th Rencontre de Blois:   
       Matter--Antimatter  Asymmetry, June 2002.}      
       }

\author{FRANS R. KLINKHAMER}

\address{Institut f\"ur Theoretische Physik, Universit\"at
         Karlsruhe, D--76128 Karlsruhe, Germany}

\maketitle\abstracts{
Electroweak baryon number violation may play a crucial role for
the creation of the matter-antimatter asymmetry in the early universe.
In this talk, we review the basic mechanism, which relies on the behavior
of chiral fermions in nontrivial Yang-Mills gauge field backgrounds.}

\section{Introduction} \label{Introduction}
The conditions for baryogenesis in the early universe are well-known
(cf. Refs.~\cite{S67,Z82}):
\vspace*{-0.25\baselineskip}
\begin{enumerate}
\item C and CP violation, \vspace*{-0.25\baselineskip}
\item thermal nonequilibrium,\vspace*{-0.25\baselineskip}
\item violation of baryon number conservation.
\end{enumerate}
\vspace*{-0.25\baselineskip}
How realistic are these requirements? Well, noninvariance under the
charge conjugation transformation (C) and the combined charge conjugation and
parity reflection transformation (CP) have been observed in the
laboratory. Also, thermal nonequilibrium can perhaps be expected
for certain (brief) epochs in the history of the early universe, as described
by the Hot Big Bang Model. But no experiment has ever seen baryon number
violation, i.e., $\Delta B \equiv B(t_\mathrm{out}) - B(t_\mathrm{in}) \neq 0$.

Strictly speaking, we know of only one physical theory that is
expected to display baryon number violation:
the \ewsm. The problem is, however, that the relevant
processes of the \sm~are only known at relatively low scattering energies,
\be
E_\mathrm{\,center-of-mass} \ll E_\mathrm{\,Sphaleron} \approx
                                M_W / \alpha \approx 10^4 \:\mathrm{GeV}  \;,
\ee
and, worse,  their cross-sections are negligible,
\be \label{tHrate}
\left. \sigma_{\Delta B\neq 0} \, \right|_{\,\mathrm{low-energy}}
\propto \exp [\,-\, 4 \pi\, \sin^2 \theta_w\, / \,\alpha\,] \approx 0\;,
\ee
with $\theta_w$ the weak mixing angle ($\sin^2 \theta_w \approx 1/4$)
and $\alpha$ the fine-structure constant ($\alpha \approx 1/137$).
Similarly, baryon-number-violating
transition rates are negligible at low temperatures, \mbox{$T \ll T_c$},
where $T_c\approx 10^2 \:\mathrm{GeV}$
sets the scale of the electro\-weak phase transition.

Clearly, we should study \ewbnv~for the conditions relevant to the early
universe---that is, for high temperatures,
\be
T \gsim 10^2 \:\mathrm{GeV}\;.
\ee
The problem is difficult but well-posed, at least within the context of the \sm.

In this contribution, we focus on the \emph{microscopic} process of \ewbnv.
This means that we must really deal with the fermions
[3--9]. 

\begin{figure}[t]
\hspace*{2.75cm}
\psfig{figure=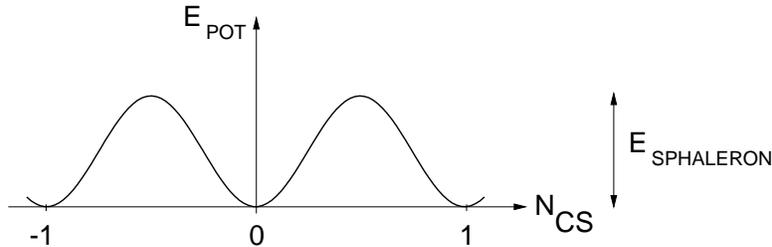,height=1.25in}
\caption{Potential energy over a slice of configuration space, parameterized by the
\CS~number.}
\end{figure}

\section{First steps}\label{Firststeps}
Consider chiral $SU(2)$ \YMH~theory with vanishing Yukawa couplings.
Actually, forget about the Higgs field, which should be reasonable
for temperatures above the electro\-weak phase transition.
Natural units with $c$ $=$ $\hbar$ $=$ $k$ $=$ $1$ are used throughout.

Now recall the existence of the
well-known triangle anomaly in the AAA-diagram,
which occurs provided the VVV-diagram is anomaly-free \cite{ABJ69}.
Here, V and A indicate vector and axialvector vertices, respectively.
(Note the obvious: the triangle anomaly is calculated with Feynman diagrams. In
other words, the calculation is \emph{perturbative}, with the
interactions ``turned off'' in the asymptotic regions of spacetime;
cf. Sec.~2 of Ref.~\cite{F49}. The
importance of this remark will become clear later on.)

The gauge vertices of the \ewsm~are V--A and the corresponding current must
be non\-anomalous (gauge invariance is needed for unitarity).
Instead, the fermion number current is anomalous \cite{H76a} and the baryon ($B$) and
lepton ($L$) number, for $N_\mathrm{fam}=3$ families of quarks and leptons,
change as follows:
\be \label{BplusLanomaly}
\Delta (B-L) = 0\;, \quad
\Delta (B+L) = 2\,N_\mathrm{fam}\; \times \;\Delta N_\mathrm{CS} \;.
\ee
In the second equation, we have on the left-hand side the difference of certain
fermion charges between the times $t_\mathrm{in}$ and  $t_\mathrm{out}$,
and on the right-hand side a characteristic of the gauge field background,
also between $t_\mathrm{in}$ and  $t_\mathrm{out}$.
Specifically, this gauge field characteristic is
\be
\Delta N_\mathrm{CS} \equiv
N_\mathrm{CS}(t_\mathrm{out}) - N_\mathrm{CS}(t_\mathrm{in})\;,
\ee
where the \CS~number $N_\mathrm{CS}(t)$ is a particular functional of the
$SU(2)$ gauge field in the temporal gauge ($A_0 =0$) at time $t$,
\be
N_\mathrm{CS}(t) = N_\mathrm{CS}[\vec A(\vec x, t)]\;.
\ee

The selection rule (\ref{BplusLanomaly}) shows that the fermion number $B+L$
changes as long as the Chern--Simons number of the gauge field changes.
But there is an \emph{energy barrier} for transitions between gauge field
vacua with different
\CS~number (Fig.~1). The top of this energy barrier corresponds to the Sphaleron
configuration, which has $N_\mathrm{CS}=1/2 \bmod 1$
(see Sec.~5 of Ref.~\cite{KM84}).

In a seminal paper \cite{H76b}, 't Hooft calculated the amplitude for
tunneling \emph{through} the barrier. For this, he
used the so-called BPST instanton, which is a finite-action solution of the
\emph{imaginary-time} \YM~theory, i.e.,  the theory in Euclidean spacetime
$(\mathrm{M},\mathrm{g})=(\mathbb{R}^4, \delta_{\mu\nu})$.
The tunneling process has then
\be \label{DeltaNCSinteger}
\Delta N_\mathrm{CS}[ A_\mathrm{\,finite\;Euclidean\;action}]  =
Q[ A_\mathrm{\,finite\;Euclidean\;action}] \in \mathbb{Z} \;,
\ee
where the topological charge $Q$
corresponds to the winding number of a particular map
\be \label{S3S3map}
\left. S^3 \right|_{|x|=\infty} \rightarrow SU(2) \sim S^3 \;.
\ee

It is important to understand this last statement. The decisive observation
is that any gauge field
with finite Euclidean action becomes \emph{pure gauge} towards infinity
($|x|^2$ $\equiv$ $x_1^2 +  x_2^2 + x_3^2 + x_4^2$ $\rightarrow$ $\infty$,
for $x_\mu \in \mathbb{R}$). Towards infinity,
the \YM~gauge field $A_\mu(x)$ can then be written as
$-\partial_\mu g\,g^{-1}$, with $g(\hat{x}) \in SU(2)$
for $\hat{x}_\mu \equiv x_\mu/|x| \in S^3$.
Hence, the gauge field at infinity is characterized by $g(\hat{x})$,
which corresponds to the map (\ref{S3S3map}).
[Note that the $SU(2)$ manifold has the topology of the three-dimensional
sphere $S^3$, because any group element $g \in SU(2)$ can be written as
$g= \vec{n}\cdot i \vec{\sigma}+ n_4\, \id$, with $|\vec{n}|^2 + n_4^2 =1$.]
The topological charge $Q$, now, measures how many times
$g(\hat{x})$ wraps around $SU(2)$ as $\hat{x}$ ranges over the 3-sphere at
infinity. This explains why the gauge-invariant
topological charge $Q$ of Eq.~(\ref{DeltaNCSinteger})
is an integer. Reference~\cite{H76a}, incidentally,  gives the selection rule
(\ref{BplusLanomaly}) in the form $\Delta (B+L) = 2\,N_\mathrm{fam}\, Q$,
at least for configurations with topological charge $Q=1$.
For later use, we prefer to write the relation (\ref{BplusLanomaly}) in terms of
$\Delta N_\mathrm{CS}$.

The property (\ref{DeltaNCSinteger}) holds only for
transitions from vacuum to vacuum, as far as the \YM~gauge field is concerned.
Practically, this means that the result can only be
relevant for processes (\ref{BplusLanomaly}) at very low energies or temperatures.
As mentioned above, the cross-section is then effectively zero
by the tunneling factor (\ref{tHrate}),
but, at least, $\Delta (B+L)$ is an \emph{integer}.

\section{Crucial question}  \label{Crucialquestion}
For \emph{real-time} processes at high energies, e.g. in Minkowski
spacetime $(\mathrm{M},\mathrm{g})=(\mathbb{R}^4, \eta_{\mu\nu})$,
the topological charge $Q$ is, in general, a \emph{noninteger}.
The reason is that the energy density of a physical \YM~gauge field
(with a conserved nonzero total energy) is never exactly zero outside
a bounded spacetime region; cf. Ref.~\cite{FKS93}. The implication is, of course,
that the expression for $\Delta (B+L)$ can no longer just have
$2\,N_\mathrm{fam}\, Q$ on the \rhs, as might be expected from the triangle
anomaly [compare with Eqs.~(\ref{BplusLanomaly}) and (\ref{DeltaNCSinteger}) above].

The question, then, is what \emph{does} appear on the \rhs,
\vspace{.25\baselineskip}
\be \label{crucialquestion}
\Delta (B+L)  \propto \mathrm{which\;gauge\;field\;characteristic\, ??}
\ee
As will be shown in Section \ref{Oldandnewresults},
the answer is fundamentally different for dissipative or
nondissipative \YM~gauge field solutions.
Here, a gauge field is called \emph{dissipative} if its energy density
approaches zero uniformly as $t\rightarrow\pm\,\infty$.

At this point, let us introduce some further terminology \cite{KL01}.
A spherically
symmetric gauge field  solution is called \emph{strongly dissipative},
if both the (3+1)-dimensional and (1+1)-dimensional energy densities
approach zero uniformly for large times ($t$ $\rightarrow$ $\pm\infty$),
and \emph{weakly dissipative},
if the (3+1)-dimensional energy density dissipates with time but not
the (1+1)-dimensional energy density.
[Note that the (1+1)-dimensional energy density divided by a factor $4\pi r^2$
corresponds to a spherically symmetric energy density in $3+1$ dimensions.]

\newcommand{\ti}{t_\mathrm{in}}   
\newcommand{\tf}{t_\mathrm{out}}

\section{Spectral flow} \label{Spectralflow}
It suffices for our calculations to consider $SU(2)$ \YMth~with a
\emph{single} isodoublet of left-handed Weyl fermions.
[A fully consistent $SU(2)$ \YMth~requires an \emph{even} number of chiral
isodoublets \cite{W82}, which is the case for the \ewsm, with
$3$ isodoublets
of left-handed quarks and $1$ isodoublet of left-handed leptons per family.
The fermion number $B+L$ of the \ewsm~follows then from the fermion number
of our simplified model by multiplication with a factor of
$(3\times 1/3 + 1\times 1)\times N_\mathrm{fam}$ $=$ $2\, N_\mathrm{fam}$.]

Start from the eigenvalue equation of the corresponding time-dependent Dirac
Hamiltonian,
\be \label{Heq}
H(\vec x, t)\,\Psi (\vec x, t)=E(t)\,\Psi (\vec x, t)\; .
\ee
Then, the resulting \emph{spectral flow}
$\mathcal{F}$ is related to the fermion number violation we are after.
The definition of spectral flow is as follows:
$\mathcal{F}[\,\tf,\ti\,]\;$ is the number of eigenvalues of the operator
considered (here, the Dirac Hamiltonian)
that cross zero from below minus the number of eigenvalues that cross
zero from above, for the time interval $[\,\ti,\tf\,]$ with $\ti < \tf$.
See Fig.~2 for a sketch and Ref.~\cite{C80} for references to the
mathematical literature.
\begin{figure}[t]
\hspace*{4.95cm}
\psfig{figure=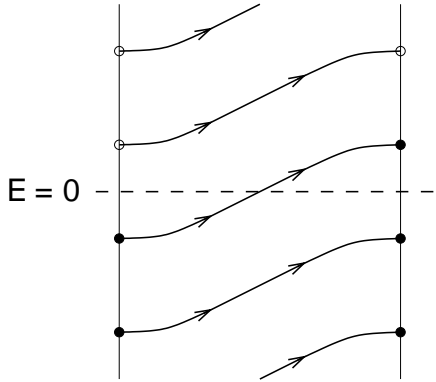,height=2in}
\caption{Spectral flow of the time-dependent Dirac
   Hamiltonian (\ref{Heq}), with $\mathcal{F}[\,\tf,\ti\,]=+1$.
   Filling the (infinite) Dirac
   sea at the initial time $\ti$ results in one extra fermion at the final time $\tf$.}
\end{figure}

\section{Old and new results on spectral flow} \label{Oldandnewresults}
For $SU(2)$ \YM~theory with a single isodoublet of chiral fermions,
the ``crucial question'' of Eq~(\ref{crucialquestion}) can be rephrased as
follows: \emph{precisely which gauge fields lead to non\-trivial spectral flow
of the Dirac eigenvalues?}

The answer is known [6--8]                  
for the case of \emph{strongly dissipative} $SU(2)$ gauge fields:
\bea \label{Fstronglydissipative}
{\cal F}[\,\tf,\ti\,] &=&      \Delta N_\mathrm{\, winding}[\,\tf,\ti\,] \no \\[0.1cm]
                      &\equiv&
            N_\mathrm{CS}[\vec{A}_\mathrm{\,associated\;vacuum}(\vec x, +\infty)] -
            N_\mathrm{CS}[\vec{A}_\mathrm{\,associated\;vacuum}(\vec x, -\infty)]\;
            ,
\eea
provided the time interval considered,
$\Delta t \equiv t_\mathrm{out}-t_\mathrm{in}$, is sufficiently large.
Here, the ``associated vacuum'' at $t=+\infty$ is the (zero-energy) vacuum
configuration which the (finite-energy) gauge field would approach starting
from $t=\tf$ and similarly for the ``associated vacuum'' at $t=-\infty$,
starting from $t=\ti$ but in the reversed direction.
The \rhs~of Eq.~(\ref{Fstronglydissipative}) is then the difference of two integers,
even though the relevant topological charge $Q$ of the gauge field may be a
noninteger.

For strongly dissipative $SU(2)$ gauge fields in the \ewsm,
the spectral flow result (\ref{Fstronglydissipative})
reproduces the selection rule (\ref{BplusLanomaly})
with $\Delta N_\mathrm{CS}$ replaced by $\Delta N_\mathrm{\, winding}$.
Note that $\Delta (B-L)$ vanishes, because the left-handed quark and lepton
isodoublets behave in the same way, namely with identical spectral flow as given by
Eq.~(\ref{Fstronglydissipative}).

Returning to the simple $SU(2)$ model with a single left-handed isodoublet,
consider next the \emph{spherically symmetric} gauge field solutions
of L\"{u}scher and Schechter (LS),
which describe collapsing and re-expanding shells of energy \cite{LS77}.
For three particular cases of these analytic solutions
(which are, in fact, ``weakly dissipative''),
the change of winding number and the spectral flow have been calculated
explicitly \cite{KL01},
\be \label{LScases}
\begin{array}{ll}
\mathrm{LS\; case\,1\; (low\; energy):}      \quad&
        \Delta N_\mathrm{\, winding}=0 \;\;\;\mathrm{and}\;\;\;  \mathcal{F}=0\,,\\[0.1cm]
\mathrm{LS\; case\,2\; (moderate\; energy):} \quad&
       \Delta N_\mathrm{\, winding}=1 \;\;\;\mathrm{and}\;\;\;   \mathcal{F}=1\,,\\[0.1cm]
\mathrm{LS\; case\,3\; (high\; energy):}     \quad&
       \Delta N_\mathrm{\, winding}=1 \;\;\;\mathrm{and}\;\;\;  \mathcal{F}=-1\,.
\end{array}
\ee
Apparently, the spectral flow need not equal the change of winding number, at
least for high enough energies with respect to a Sphaleron-like barrier of the
potential energy.
In other words, the previous result (\ref{Fstronglydissipative})
does not hold in general.

The correct relation for the spectral flow in a generic spherically symmetric
$SU(2)$ gauge field background follows from the existence of
\emph{another} gauge field characteristic, $\Delta N_\mathrm{\, twist}$,
so that \cite{KL01}
\be \label{newselectionrule}
\left[\,\mathcal{F} = \Delta N_\mathrm{\, winding} +
\Delta N_\mathrm{\, twist}\,\right]_\mathrm{\,spherically\;symmetric\;fields} \,.
\ee
Here, $N_\mathrm{\, twist}(t)$ is an integer number which can be calculated
directly from the $SU(2)$ gauge field configuration at time $t$.
For the special cases  considered in Eq.~(\ref{LScases}), the relation
(\ref{newselectionrule}) is verified with
\be \label{DeltaNtwist}
\begin{array}{rcll}
\Delta N_\mathrm{\, twist}&=&0\,,  &\quad \mathrm{for\;LS\; case\;1\;and\;2\,,}\\[0.1cm]
\Delta N_\mathrm{\, twist}&=&-2\,, &\quad \mathrm{for\;LS\; case\;3\,.}
\end{array}
\ee
It should be emphasized that the new selection rule (\ref{newselectionrule})
has two integers on the
\rhs, whereas the topological charge $Q$ may be a noninteger. Indeed,
the LS cases 2 and 3 have $Q=0.70$ and $Q=0.13$, respectively.

For \emph{weakly dissipative} or \emph{nondissipative} $SU(2)$
\YM~gauge fields in the \ewsm, one has thus
\bea  \label{newBplusLanomaly}
\Delta (B-L) &=& 0\;,\nonumber \\
\Delta (B+L) &=& 2\,N_\mathrm{fam}\times
\Big(\Delta N_\mathrm{\, winding}
+ \underline{\mathrm{extra\;terms}}\,\Big)\;,
\eea
with $\Delta N_\mathrm{\, winding}$ as defined by the
\rhs~of Eq.~(\ref{Fstronglydissipative}).
According to Eq.~(\ref{newselectionrule}), there is a single ``extra term''
for the case of spherically symmetric fields, namely $\Delta N_\mathrm{\, twist}$.
But, in general, the ``extra terms'' of Eq.~(\ref{newBplusLanomaly}) are not
known. Note also that the issue of gauge invariance deserves particular care.

\section{Summary} \label{Summary}
To our knowledge, there is only one established theory in elementary particle
physics for which \bnv~can
be expected to occur, namely the \ewsm.
(Evaporating black holes may or may not violate baryon number
conservation. Our current understanding does not allow for definitive
statements about the ultimate fate of black holes; cf. Ref.~\cite{H92}.)

Most discussions of \ew~baryogenesis (cf. Ref.~\cite{RS96})
have been based on the
selection rule (\ref{BplusLanomaly}), which holds in particular
for the tunneling process at low energies \cite{H76a,H76b}.
As remarked in the second paragraph of Section~\ref{Firststeps}, this
relation has first been derived from perturbation theory \cite{ABJ69,F49},
with the interactions in the asymptotic regions of spacetime ``turned-off.''
It is then not altogether surprising that we have found relation
(\ref{BplusLanomaly}) to be invalid for high-energy gauge field backgrounds
which are weakly dissipative or nondissipative \cite{KL01}.
Of course, precisely these fields are relevant to the physics of the early universe.

At this moment, we have only a partial result  for the correct
selection rule, namely Eqs.~(\ref{newselectionrule}) and (\ref{newBplusLanomaly})
for the case of \emph{spherically symmetric} \YM~gauge fields.
To generalize this result to \emph{arbitrary} \YM~gauge fields will be difficult,
but is absolutely necessary for a serious discussion of \ewbnv~in
the early universe.

\section*{Acknowledgments}
The author would like to thank the organizers of this conference
for a splendid meeting and J.~Schimmel for help with the figures.

\end{document}